\documentclass[11pt]{article}
 \pdfoutput=1
\usepackage{graphicx}
\usepackage{savesym}
\usepackage{amsmath}
\savesymbol{iint}
\usepackage{txfonts}
\restoresymbol{TXF}{iint}

\usepackage{amssymb,amsfonts,verbatim}

\usepackage{natbib}
\usepackage[hang,nooneline]{subfigure}
\usepackage{aas_macros}
\usepackage{amsmath}
\usepackage{url}
\newcommand{\deleteopen}{{\bf{To delete[}}\marginpar{$\Downarrow$}}
\newcommand{\deleteclose}{{\bf{]To delete}}\marginpar{$\Uparrow$}}
\begin{document}
%%%%%%%%%%%%%%%%
\title{The SPOCA-suite: a software for extraction and tracking of Active Regions and Coronal Holes on EUV images}

 \author{V. Delouille$^{1,2}$ \and B. Mampaey$^{1}$ \and C. Verbeeck$^{1}$ \and R. de Visscher$^{1}$}
      \footnotetext[1]{Royal Observatory of Belgium, Circular Avenue 3, B-1180 Brussels, Belgium}
      \footnotetext[2]{Corresponding author. Email: v.delouille@sidc.be}
 
\date{\today}

% \abstract{}{}{}{}{} 
% 5 {} token are mandatory
 \maketitle
 
\begin{abstract}
 
   Precise localisation and characterization of active regions and coronal holes as observed by EUV imagers are crucial for a wide range of solar and helio-physics studies.
   We describe a segmentation procedure, the \emph{SPOCA-suite}, that produces catalogs of Active Regions and Coronal Holes on SDO-AIA images.
   The method builds upon our previous work on \lq Spatial Possibilistic Clustering Algorithm' (SPOCA) and substantially improve it in several ways.
   The SPOCA-suite is applied in near real time on AIA archive and produces entries into the AR and CH catalogs of the Heliophysics Event Knowledgebase (HEK) every four hours.We give an illustration of the use of SPOCA for determination of the CH filling factors.
 This reports is intended as a reference guide for the users of SPoCA output.
   \end{abstract}
   %%%%%%%%%%%%%%%%%%%%%%%%%%%%%%% BEGIN COMMENT
   
\vspace{0.5cm}
\vspace{0.5cm}
  
 \section{Introduction} 
 
 Accurate determination of Active Regions and Coronal Holes properties on coronal images is important for a wide range of applications. As regions of locally increased magnetic flux, the active regions  (AR) are the main source of solar flares. A catalog of ARs describing key parameters such as their
 location, shape, area, mean and integrated intensity would allow for example to relate
 those properties to the occurence of flares.  Having a bounding box of ARs can prove useful when studying several thousands of ARs together e.g. to make statistical analysis such as oscillations of loops.
  Precise localisation of coronal holes on the other hand is important because of the strong association between coronal holes and high-speed solar wind streams~\citep{1973SoPh...29..505K}.
Finally,  solar EUV flux plays a major role in Solar-Terrestrial relationships, and hence  an accurate monitoring of coronal holes (CH), quiet sun (QS) and active regions (AR) is desirable as input into solar EUV flux models. 
 
\begin{figure*}[t]
\vspace*{2mm}
\begin{center} 
 \includegraphics[height=8cm]{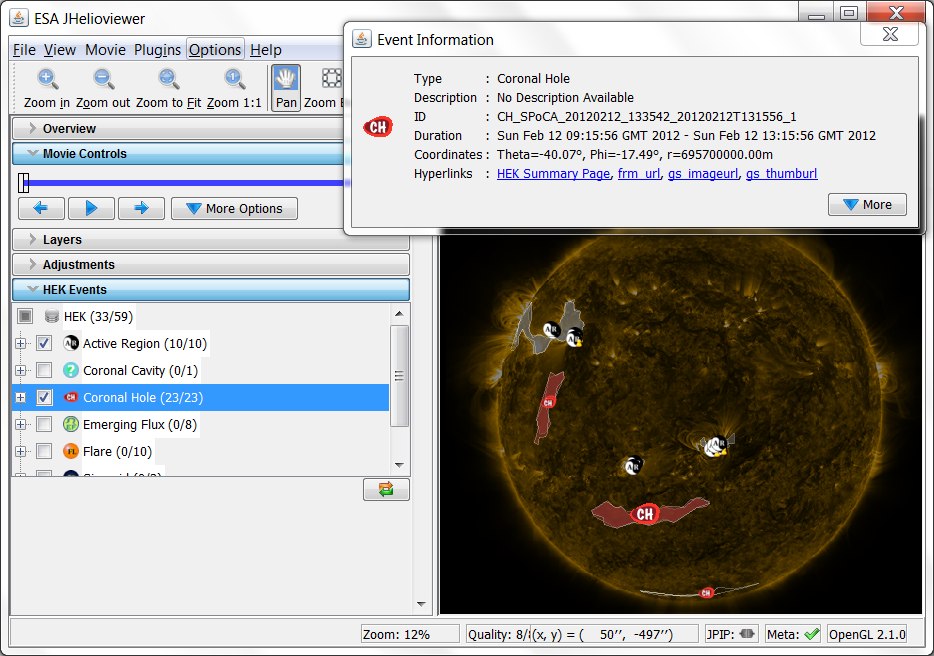}
 \caption{\label{F:JHV}  Screenshot from the ESA JHelioviewer tool. The picture on the right displays the AIA 171Å image taken on 12 February 2012 at 9:02:12 together with Active Region and Coronal Hole location and chain-code information that are recorded in the HEK. An 'Event Information' window pops up when clicking on an event or feature (here the large Coronal Hole located in the South hemisphere)}
    \end{center}
 \end{figure*}

In this paper, we present the SPOCA-suite, a set of algorithms that allows separating and extracting CH, AR, and QS on EUV images. We detail the parameterization needed to treat SDO-AIA images. The SPOCA-suite is applied in near real time on AIA archive and produces entries into the AR and CH catalogs of the Heliophysics Event Knowledgebase every four hours. The HEK is further linked through an API to the  graphical interface called \emph{isolsearch}~\footnote{See \url{http://www.lmsal.com/hek/hek_isolsearch.html}}, to the \emph{ontology} software package of \emph{Solarsoft}~\footnote{See Section 5.2 of \url{https://www.lmsal.com/sdodocs/doc/dcur/SDOD0060.zip/zip/entry/index.html}}, and to the JHelioviewer visualization tool~\footnote{See \url{jhelioviewer.org}}. The code is written is C++, with some wrapper in IDL and Python. It is available upon request to the corresponding author. Figure~\ref{F:JHV} shows a screenshot from the ESA JHelioviewer tool with CH and AR boundaries overlays.

This paper is intended as the reference guide for the users of the SPOCA output. It presents the improvements made with respect to the   the  previous work on the Spatial Possibilistic Clustering Algorithm~\citep{2009A&A...505..361B}. At the core  of the SPOCA-suite lies a multi-channel unsupervised fuzzy clustering method that segments EUV images into different regions according to their intensity level.

 %%%%%%%%%%%%%%%%%%%%%%%%%%%%%%%%%%%%%%%%%%%%%%%%%%%%%%%%%%%%%%%%%%
 %{\bf Brief review of literature, to be shortened and updated}
 
 Development of automated solar feature detection and identification methods has increased dramatically in recent years due to the growing volume of data available. An overview of the fundamental image-processing techniques used in these algorithms is presented in \citet{Aschwanden_2010}. These techniques are tailored to detect features in various types of observations at different heights in the solar atmosphere, see for example~\citep{2012SoPh..275...79M,PerezSuarez2011}.
 For example, regions of locally intense magnetic flux are observed as dark spots (sunspot) in white light, CaII, or continuum images, as an interlace of positive and negative magnetic field value  (magnetic AR) in magnetograms, and as patches of enhanced brightness in chromospheric (plages) and coronal images (AR).

 Image segmentation methods are typically classified into three broad categories:  region-based methods, edge-based methods and hybrid approaches.
 Region-based methods seek a partition of the image  satisfying an homogeneity criterion (on mono-,multispectral gray levels or higher level attributes such as texture or feature vectors modeling pixels and their neighborhood).   The dual edge-based approaches aim at characterising image discontinuities, and thus locating region boundaries. Primal edge-based methods seek for maximum of intensity gradients, using either spatial or frequency filters, or zeros in the Laplacian of the image, often pre processed by a low pass Gaussian filtering, due to the Laplacian sensitivity to noise. Finally, the hybrid methods either consider a cooperation between region and contour approaches, or process some other original method.
 
 Region-based methods for the detection of sunspots include thresholding against background~\citep{Pettauer97,Colak2008}, histogram-based thresholding \citep{Steinegger97}, region-growing method~\citep{Preminger97}, or bayesian approach~\citep{2002ApJ...568..396T}.   \citep{Colak2008,Nguyen2005}  combine thresholding and machine learning techniques to extract and classify sunspots according to the McIntosh system. \citep{Curto2008,Watson2009} both use an edge-based approach based on mathematical morphology. \cite{Zharkov2004} employ a hybrid method that combines edge-based and thresholding approaches, whereas \cite{Lefebvre2004} presents a singular spectrum analysis to detect sunspots and faculae at the solar limb.
  
Active regions as observed by magnetograms can be extracted and characterized by means of region growing techniques~\citep{Benkhalil2003, 2005SoPh..228...55M, Higgins2010}, thresholding in the intensity~\citep{Qahwaji2006,2009SpWea...706001C} or wavelet domain~\citep{2010ApJ...717..995K}.
\cite{2011SoPh..tmp..369V} provides a detailed comparaison of outputs from sunspots, magnetic, and coronal active regions using one month of SOHO-EIT data. At the chromospheric level, network and plage regions are separated using thresholding methods~\citep{Steinegger98,Worden99}, possibly combined with region-growing techniques~\citep{2006SoPh..235...87B}. Coronal active regions are segmented using either local thresholding and region-growing method~\citep{2006SoPh..235...87B}, supervised~\citep{Dudok06,2011SoPh..tmp..388C}, or unsupervised~\citep{2009A&A...505..361B} classification. \cite{2005SoPh..228...43R} compares segmentation results of pixelwise fractal dimension of EIT images using thresholding, region-growing technique, and supervised classification.

Coronal holes are regions of lower electron density and temperature than the typical quiet Sun, and appear thus as dark regions in EUV and X-ray images. 
However, automated detection of coronal holes based on intensity thresholding in one wavelength (EIT 28.4nm wavelength in~\citep{2009SoPh..260...43A,2009SoPh..260..191O} or Soft X-ray images in~\citep{2007SoPh..240..315V,2011A&A...526A..20V}) is intrinsically complicated due to the presence
of filaments and transient dimmings of same intensity level. 
%----
To resolve this ambiguity, it is necessary to make use of additional information coming from magnetograms, from other wavelengths, or from time evolution of the feature in order to check the consistency of a CH candidate  with actual physical parameters.
For example, \cite{2005ASPC..346..261H} first use a fixed thresholding based on two-day average He I 1083 nm spectroheliograms and thereafter check the unipolarity of the CH candidates using photospheric magnetograms.  \cite{2005ASPC..346..251T} use a combination of fixed thresholdings on multiple wavelengths (the four SOHO-EIT bandpasses, He I 1083, magetograms, and H$\alpha$ images) to determine stringent criteria for a region to belong to a CH, whereas ~\cite{2011SoPh..274..195D} use a similar technique on synoptic maps.
The approach in~\citep{2008SoPh..248..425S} is to first perform histogram equalization and fixed thresholding to extract low intensity regions on the four bandpasses of SOHO-EIT. In a second stage  statistics on magnetic field parameters measured by SOHO-MDI are evaluated to discriminate between filaments and coronal holes. A similar methodology is used in~\cite{2009SoPh..256...87K} with the difference that it first  detects  low intensity regions using local histogram of SOHO-EIT, STEREO-EUVI, and Hinode-XRT images.
Other methods include a watershed approach~\citep{Nieniewski02}, perimeter tracing for polar coronal hole using morphological transform and thresholding 
\citep{2009SoPh..257...99K}, and the use of imaging spectroscopy to separate quiet sun from coronal hole emission~\citep{2005SoPh..226....3M}.
%-----
Finally, the classification approach in~\citep{Dudok06,2011SoPh..tmp..388C,2009A&A...505..361B} allows  separating both active regions and coronal holes using brightness intensity observed in one or in multiple bandpasses.

%------- End Bibliography...

In this paper, we describe the improvements made with respect to the paper in~\citep{2009A&A...505..361B} in terms of robustness and stability. 
For example, the segmentation in~\cite{2009A&A...505..361B} required the pre-determination of centers of classes before applying it on a large data set, which is difficult to implement when the segmentation is performed continuously on a stream of data.

As part of the Feature Finding Team~\citep{Martens_etal_2011}, programs have been specifically written to run in near real time on SDO-AIA images. It results in two separate modules: SPOCA-AR module for Active Regions, and  SPOCA-CH for coronal.

Section~\ref{S:spoca} below describes the algorithm, and Section~\ref{S:aia} provides some illustration of the algorithm on SDO-AIA data.

%-------------------------------------------------------------
\section{The SPOCA-suite}
\label{S:spoca}
The SPOCA-suite implements  three types of fuzzy clustering algorithms: the Fuzzy C-means (FCM), a regularized version of FCM called Possiblistic C-means (PCM) algorithm, and a Spatial Possiblistic Clustering Algorithm (SPOCA) that integrates neighboring intensity values.

The description of the segmentation process in terms of fuzzy logic was motivated by the facts that information provided by an EUV solar image is noisy and subject to both observational biases (line-of-sight integration of a transparent volume) and interpretation (the apparent boundary between regions is a matter of convention). Fuzzy measures are able to represent ill-defined classes (without a clear-cut boundary) in a natural way.

Let $x_j\in\mathbb{R}^p$ be a $p$-dimensional feature vector that describes the Sun at a particular location. In our case, $x_j$ is a $N-$dimensional vector of intensities recorded in $p$ different channels. A fuzzy clustering algorithm searches for $C$ compact clusters amongst  the $x_j$'s in $X=\lbrace x_j\rbrace$. It does so   by computing both a fuzzy partition matrix
$U=(u_{ij}),1\leq i\leq C, 1\leq j\leq N$ and the cluster centers $B=(b_i\in\mathbb{R}^p,1\leq i\leq C)$. The scalar  
$u_{i,j}=u_i({x_j})\in[0,1]$  is called the membership degree of ${x_j}$ to class $i$.

We now describe  the FCM and PCM algorithm as they are used in the SPOCA-suite. A description of the Spatial Possibilistic clustering algorithm is available in~\cite{2009A&A...505..361B}.

\subsection{Fuzzy Clustering Algorithm}
Since its introduction by Bezdek \citep{Bezdekbook}, the Fuzzy C-Means (FCM) algorithm was widely used in pattern recognition and image segmentation, in various fields including medical imaging \citep{Philipps95,Bezdek97}, remote sensing images \citep{Rangsanseri98,Melgani00}, color image segmentation in vision \citep{Baker03}. 
As compared to the crisp (traditional) method, the fuzzy segmentation more often reaches a global optimal, rather than merely a local optimum~\citep{Trauwaert91}.

The idea behind FCM is the minimisation of the total intracluster variance: 
\begin{equation}
\label{E:JFCM}
J_{FCM}(B,U,X) = \displaystyle\sum_{i=1}^C\displaystyle\sum_{j=1}^Nu_{ij}^md^2({x_j},{b_i})
\end{equation}
subject to
\begin{equation}
(\forall i\in\{1\cdots C\})\displaystyle\sum_{j=1}^N u_{ij}<N \quad\textrm{ and }\quad 
(\forall j\in\{1\cdots N\})\displaystyle\sum_{i=1}^C u_{ij}=1 \label{probaconstraint}
\end{equation}
where $d$ is a metric in $\mathbb{R}^p$ (typically, the Euclidian distance), and $m$ is a parameter that 
controls the degree of fuzzification ($m=1$ means no fuzziness). In practice, a value of $m=2$ is often chosen, as it allows for a fast computation in the iterative scheme. In our application we consider either one ($p=1$) or two channels ($p=2$).

The minimization of~(\ref{E:JFCM}) is reached when
\begin{equation}
\label{E:computeUB}
u_{ij} =  \frac{1}{\sum_{k=1}^C \left(\frac{d^2(x_j,b_i)}{d^2(x_j,b_k)} \right)^{1/(m-1)} }\quad\textrm{ and }\quad 
b_i = \frac{\sum_{k=1}^N u_{ik}^m x_k}{\sum_{k=1}^N u_{ik}^m}
\end{equation}
The values satisfying~(\ref{E:computeUB}) are obtained through an iterative procedure. 

The shortcomings of FCM are of two types. First, it is sensitive to noise and outliers~\citep{Krishnapuram93}. 
Second, it is theoretically not satisfying since the value of one center in~(\ref{E:computeUB}) depends on the value of the other centers.

In practice however FCM provides the best results for extracting Coronal Holes out of the (almost noise free) 19.3nm AIA images. Given the size of AIA images, FCM is applied on the histogram of normalized intensity values, rather than on individual values. The normalization consists in dividing by exposure time, correcting for limb brightness enhancement, and dividing by the median value (see Section~\ref{S:pre-processing}). A binsize of 0.01 for the histogram provides the same numerical precision as if individual values were used.

\subsection{Possibilistic C-means algorithm}
To obtain a formulation for $u_{ij}$ that depends only on the distance to the center of class $i$, \cite{Krishnapuram93,Krishnapuram96} propose to minimize the objective function 
\begin{equation}
J_{PCM}(B,U,X) = \sum_{i=1}^C \left(\sum_{j=1}^Nu_{ij}^md^2({x_j},{b_i}) + \eta_i 
\sum_{j=1}^N (1-u_{ij})^m\right )  \label{E:objective_fct}
\end{equation}
subject to
\begin{equation}
(\forall i\in\{1\cdots C\})\displaystyle\sum_{j=1}^Nu_{ij}<N \quad \mbox{and} \quad
(\forall j\in\{1\cdots N\})\displaystyle\max_i u_{ij}>0\label{constraintmax}
\end{equation}
The first term  of $J_{PCM}$ in~(\ref{E:objective_fct}) is the intracluster variance, whereas the second term, stemming from the relaxation of the probabilistic constraint in (\ref {probaconstraint}), enforces $u_{ij}$ to depend only on $d({x_j},{b_i})$.

Parameter $\eta_i$ in~(\ref{E:objective_fct}) is homogeneous to a squared distance. It can be fixed, or updated at each iterations.
 \cite{Krishnapuram93} proposed to compute $\eta_i$ as the intra-class dispersion: 
\begin{equation}
\label{E:eta}
\eta_i=\frac{\displaystyle\sum_{j=1}^Nu_{ij}^m d^2({x_j},{b_i})}{\displaystyle\sum_{j=1}^Nu_{ij}^m}
\end{equation}
The solution of the minimization of~(\ref{E:objective_fct}) satisfies:
\begin{eqnarray}
\label{E:u_ij}
u_{ij}&=&\left[1+\left(d^2({x_j},{b_i})/\eta_i \right)^{\frac{1}{m-1}} \right]^{-1}\label{E:uijPCM}\\
b_{i}&=&\displaystyle\sum_{j=1}^N u_{ij}^m  {x_j}/\displaystyle\sum_{j=1}^N u_{ij}^m \label{E:biPCM}
\end{eqnarray}

In practice, PCM is initialized by one iteration of FCM, which allows for computation of $\eta_i$ as in~(\ref{E:eta}). \cite{Krishnapuram93} proved the convergence of iteration (\ref{E:uijPCM})-(\ref{E:biPCM}) for fixed $\eta$.  PCM is more robust to noise and outliers than FCM, and provides independent functions $\mathbf{u_i}=\{u_{ij}, j=1,\ldots, N\}$.
It must be corrected however from coincident clustering (Section~\ref{S:coincident_clustering}), and a proper choice of the parameter $\eta$ must be made (Section~\ref{S:constraint_etai}). The SPOCA-AR module of the HEK uses this corrected PCM algorithm on the AIA 17.1nm and 19.3nm bandpasses. Similar to the SPOCA-CH module, it is applied on histogram intensity values rather than on individual pixel intensity values.

\subsubsection{Coincident Clustering}
\label{S:coincident_clustering}

\begin{figure*}[htbp]
 \vspace*{2mm}
 \begin{center}
 \includegraphics[width=6cm]{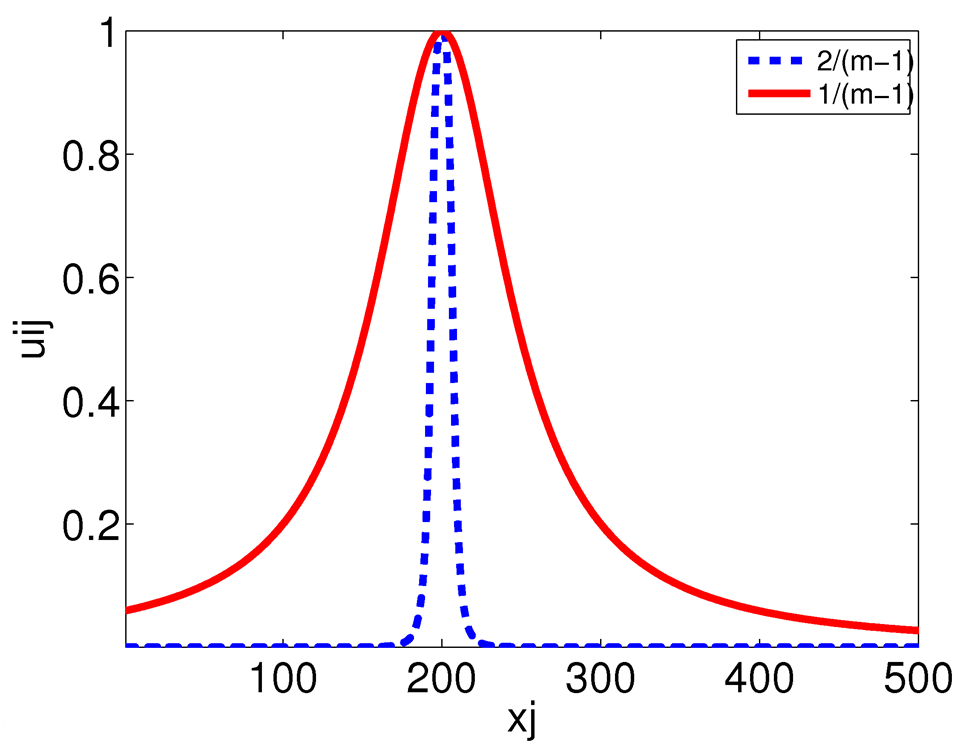}
 \caption{ \label{F:coincident} Comparaison of membership functions $u_{ij}$ when the exponent is chosen equal to $1/(m-1)$ (red line) and $2/(m-1)$ (blue dashed line). The blue dashed line is more compact, which will lead to the choice of distinct centres of classes.}
 \end{center}
 \end{figure*}

The original PCM suffers from convergence to a unique center. This is a typical feature of possibilistic clustering algorithms called \lq coincident clustering'~\citep{Krishnapuram96}. To circumvent this problem we use membership functions $u_{ij}$ which are more compact and hence do not overlap so easily.
More precisely,  exponent $1/(m-1)$ in~(\ref{E:uijPCM}) is replaced  by $2/(m-1)$, see Figure~\ref{F:coincident} for a graphical representation. We call PCM2 the algorithm where the exponent in the membership function $u_{ij}$ is taken equal to $2/(m-1)$.

\subsubsection{Constraints on the parameter $\eta_i$}
\label{S:constraint_etai}
The dynamical range of intensities differs amongst the CH, QS, and AR classes, with active regions having the largest spread in intensities. The parameter  $\eta_i$ as computed in~(\ref{E:eta}) can be viewed as a measure of dispersion or variance within a class. In case $\eta_{AR}$ becomes prohibitively large, the value of $u_{AR,j}$ for dark pixels $x_j$ can be higher than $u_{QS,j}$ or $u_{CH,j}$, as illustrated in Figure~\ref{F:eta-condition}(a).  To avoid this situation we must enforce the following inequalities, as derived in Appendix~\ref{A:constraint_eta}:

\begin{equation}
\label{eta-condition}
\frac{\eta_{QS}}{\eta_{CH}} < \frac{b_{QS,q}}{b_{CH,q}},
\frac{\eta_{AR}}{\eta_{CH}} < \frac{b_{AR,q}}{b_{CH,q}},
\frac{\eta_{AR}}{\eta_{QS}} < \frac{b_{AR,q}}{b_{QS,q}}, \mbox{ for } q = 1, \dots, p.
\end{equation}
with $b_{CH,q} , b_{QS,q} ,$ and  $b_{AR,q}$ the center of classes for the q-th channel and for CH, QS and AR respectively.
Figure~\ref{F:eta-condition} shows an example of segmentation with and without constraints on $\eta_i$. Without constraints, some CH areas gets classified in the same class as the AR. This problem is solved when constraints~(\ref{eta-condition}) are introduced.

For solar EUV images, the combination of the iteration scheme (\ref{E:eta})-(\ref{E:biPCM}) tends to produce $\eta_{CH}$-values that converge to zero. Due to the 
condition~(\ref{eta-condition}) on $\eta_{QS}$ and $\eta_{AR}$, these two parameters are dragged along to converge to zero as well.  Our iterative scheme therefore freezes the value of $\eta_i$ when it has changed by a factor $\alpha$ with respect to its starting value. In other words, formula~(\ref{E:eta}) is used until iteration ${it}$ is reached with:
\[
\eta_i^{it} / \eta_i^1 > \alpha {\rm\ or\ } \eta_i^1 / \eta_i^{it} > \alpha~.
\]
For the next iterations, we keep $\eta_i^{it}$ fixed. Satisfactory results have been obtained on a variety of datasets and instruments with $\alpha = 100$.

\subsection{Smooth variation of center values}
In order to have a smooth variation of the center values over time, the centers chosen at time $t$ in the HEK are in fact the median of the last 10 centers obtained at previous time stamp $t,t-1, \ldots, t-9$. 
In order to get the membership map corresponding to this smooth value of the center $b_i$ an attribution procedure using left part of formula~(\ref{E:computeUB}) (for FCM), or formula~(\ref{E:uijPCM})(for PCM) is performed.

%----------- Value of parameters used for the HEK
% SPOCA_AR
%instrument = 'AIA'
%
%spocaArgs = ['--preprocessingSteps', 'DivExpTime,ALC,DivMedian', $
%			'--classifierType', 'HPCM2', $
%			'--numberClasses', '4', $
%			'--precision', '0.0015', $
%			'--radiusratio', '1.2', $
%			'--binSize', '0.01,0.01', $
%			'--segmentation', 'threshold', '-t', '2,0,0.0001', $
%			'--numberPreviousCenters', '10', $
%			'--regionStatsPreprocessing', 'NAR,DivExpTime', $ (why to divide again by ExpTime?)
%			'--regionStatsRadiusRatio', '0.95', $
%			'--maps', 'A']
%
%chaincodeArgs = ['--chaincodeMaxPoints', '100', $
%			'--chaincodeMaxDeviation','10' ]
%
%trackingArgs = ['--max_delta_t', '14400'] ; == 4h
%
%trackingOverlap = 2
%
%w171='171'
%w195='193'
%-----------------------------------------------
% SPOCA_CH
%instrument = 'AIA'
%
%spocaArgs = ['--preprocessingSteps', 'DivExpTime,ALC,ThrMax80,TakeSqrt', $
%			'--classifierType', 'HFCM', $
%			'--numberClasses', '4', $
%			'--precision', '0.0015', $
%			'--radiusratio', '1.2', $
%			'--binSize', '0.01', $
%			'--segmentation', 'max', $
%			'--numberPreviousCenters', '10', $
%			'--regionStatsPreprocessing', 'NAR,DivExpTime', $
%			'--regionStatsRadiusRatio', '0.95', $
%			'--maps', 'C']
%
%chaincodeArgs = ['--chaincodeMaxPoints', '100', $
%			'--chaincodeMaxDeviation','10' ]
%
%trackingArgs = ['--max_delta_t', '14400'] ; == 4h
%
%trackingOverlap = 2
%
%w195='193'

\subsection{Segmented maps}
Given the membership maps $U$ and centers $B$ a segmention map can be obtained using various decision rules

\begin{description}
\item[\bf{Maximum}:]
The most common rule is to attribute  a pixel $j$ to the class $c$ for which it has the maximum membership value:  $\{u_{c,j} = \max_{i\in\{{1,\ldots, C}\}} u_{ij}\}$.  This rule is used in the SPOCA-CH module.
\item[\bf{Threshold}]
On EUV images the QS class contains typically the largest number of pixels, and hence the center of QS class is the most stable over time. On the contrary, the cardinality of points belonging to  the AR class varies a lot over time, resulting in a high variation of center of AR class. In order to have a stable segmentation of AR over time, the SPOCA-AR module therefore uses a threshold on the QS membership class to define the belonginess of a pixel to the AR class. All pixels $j$ which are higher than $b_{QS}$ and for which $u_{QS,i}$ is lower than 0.0001 are attributed to the AR class.
\item[\bf{Closest}]
This rule attributes a pixel to the class for which the Euclidian distance to the center class is the smallest. 
\item[\bf{Merge}]
This more complex procedure uses sur-segmentation and merging of classes and has been described in~\citep{2009A&A...505..361B}.
\end{description}

\subsection{Pre-processing}
\label{S:pre-processing}
Some pre-processing steps are needed in order to obtain an accurate segmentation of the images.

First, images are calibrated using \emph{solar software} routines and intensities are divided by the exposure time. 

Second, the SPOCA-suite might be applied on linear or on square root transformed images. A square root transform is akin to an Anscombe transform~\citep{ansc} which has the property of approximately converting Poisson noise into Gaussian noise. This is especially useful for the extraction of low-intensity regions such as coronal holes which are affected by Poisson noise.

Third, the limb brightening effect has to be corrected before any segmentation based on intensity can be applied. A first correction was proposed in~\citep{2009A&A...505..361B}. It consists of applying a 
 a polar transform to represent the image $I$ in a $(\rho,\theta)$ plane, with origin at the solar disc center. The polar transform is a conformal mapping from points in the cartesian plane $(x,y)$ to points in this polar plane, described by: $\rho=\sqrt{x^2+y^2},\theta=atan(y/x)$. The intensity is specified as a function of $\rho$ using the integral $g(\rho)=\int_0^{2\pi}I(\rho,\theta)d\theta$. Denoting $m_\odot$ the median value of intensities on the on-disc part of the Sun, the image $I_{corr}$ corrected for the enhanced brightness at the limb is computed as:
\begin{equation}
 I_{corr}(\rho,\theta) = m_\odot\frac{I(\rho,\theta)}{g(\rho)}~,
\label{E:Icorr}
\end{equation}
for values of $\rho$ ranging between 95\% and 105\% $R_\odot$. $I_{corr}$ is then finally remapped to the cartesian plane. 

Such abrupt correction leads to disconintuities in the images around these radial distances, as can be seen on the images in Figure~\ref{F:eta-condition}. We propose instead to apply a correction $I_{smooth}$:
\begin{equation}
I_{smooth}(x,y) = \left(1 - f(\rho(x,y))\right) I(x,y) + f(\rho(x,y)) I_{corr}(x,y).
\end{equation}
where $f(\rho(x,y))$ introduces a smooth transition between the zones not corrected (when $\rho$ is between 0 and  $r_1$ or above $r_4$)  and the zones $\rho \in [r_2,r3]$ that are fully corrected using equation~(\ref{E:Icorr}):
\[
f(\rho) = \left\{
\begin{array}{lll}
0 &\mbox{if}&  \rho \leq r_1 {\rm\ or\ } \rho \geq r_4 \\
1 &\mbox{if}& \rho \in [r_2,r_3]\\
 \frac{1}{2}\sin\Big(\frac{\pi}{r_2 - r_1} (\rho - \frac{r_1 + r_2  }{2})\Big) + \frac{1}{2}
&\mbox{if}&  r_1 \leq \rho \leq r_2 \\
\frac{1}{2}\sin\Big(\frac{\pi}{r_4 - r_3} (\rho + \frac{r_4 - 3 r_3}{2})\Big) + \frac{1}{2}  &\mbox{if} &r_3 \leq \rho \leq r_4
\end{array}
\right.
\] 
A simulation study was conducted to determine the optimal values of $[r_1, r_2, r_3, r_4]$ for a given instrument.
The mean square error (MSE) between corrected limb values and the on-disc median intensity is used as criterion.
For SDO-AIA, the  values that minimizes the MSE are (in \% of $R_\odot$) are $[70, 95, 108, 112]$ .
Fnally, $I_{smooth}$ is divided by its median value.

%% Note : in the SPOCA program, these values are in constants.h

\subsection{Region extraction and post-processing}

To extract regions (that will be labelled as Active Regions or as Coronal Holes) from segmented maps, the following post-processing steps are implemented:
\begin{enumerate}
\item Compute a sinusoidal projection map~\citep{projections}. This improves the determination of regions towards the limb. 
\item Clean the segmented map by removing elements smaller than 6~\emph{arcsec} using a morphological erosion
\item Aggregate neighboring blobs by performing a morphological closing, which consists of a dilatation by 32 arcsec followed by an erosion.
\item Compute the inverse of sinusoidal projection
\end{enumerate}

The reader is refered to~\cite{2009A&A...505..361B} for an introduction to mathematical morphology in this context. The equirectangular and Lambert cylindrical projections are also implemented in the SPOCA-suite. Our test on SDO data shows that the sinusoidal projection gives the best results.

To remove spurious regions,  a final cleaning is performed as follows. Active regions smaller than $1500~\mbox{arcsec}^2$ and coronal holes smaller than $3000  \mbox{arcsec}^2$ are discarded.

Moment statistics and properties such as area, barycenter location, bounding box are then computed\footnote{The list of all features computed can be found in \url{http://www.lmsal.com/hek/VOEvent_Spec.html}}.

A coronal hole has relatively smooth boundaries. Having an accurate estimation of its shape and localisation is crucial for space weather purpose, since coronal holes are most geoeffective when they are located at the central meridian and near the equator.
We computed a chain code for the coronal holes using per default a maximum of 100 points to describe the contour. The details of the algorithm are described in Appendix~\ref{A:chain_code}. Similarly, we provided a chain-code for Active Regions. Note however that in order to represent in an optimal way the non-smooth AR boundaries more than 100 points would be needed. 

\subsection{Tracking over time}
% once you have found the connected components: look in two successive images and see if CC at time t and time t+1 overlap.

Following the aggregation of regions described in the previous section, an active region is defined as a coherent group of corresponding active region blobs, and similarly for the coronal holes. 
The goal of tracking is to appoint the same ID number to a physical region (AR or CH) over time. A region observed at timestamp $t$ can follow the region observed at previous timestamp, but it can also split (and produce two children),  or it can merge with other neighboring regions. 

Our tracking scheme amounts to create a directed graph $(N,E)$ where $N$ is the list of nodes representing individual regions and $E$ is the list of edges between regions. 
An edge between a region observed at time $t_1$ and another observed at time $t_2$ is created if their time difference $t_2-t_1$ is smaller than some value, if they overlay, and if there is not already a path between them. 
It is possible to first derotate the regions maps before comparing them. This is necessary when large time difference are involved.

Coronal holes are long-lived features. We can include this information in the tracking and keep only coronal holes that are older than three days. Our analysis of SDO images during the year 2011 shows that a coronal hole candidate detected for more than three consecutive days exhibits the expected magnetic properties characteristics of unipolar regions. 
Hence we report to the HEK only the CHs that are older than three days.

\section{Analysis of SDO-AIA images}
\label{S:aia}
\begin{figure*}[htpb]
\vspace*{2mm}
\begin{center}

\begin{tabular}{cc}
\includegraphics[height=8cm]{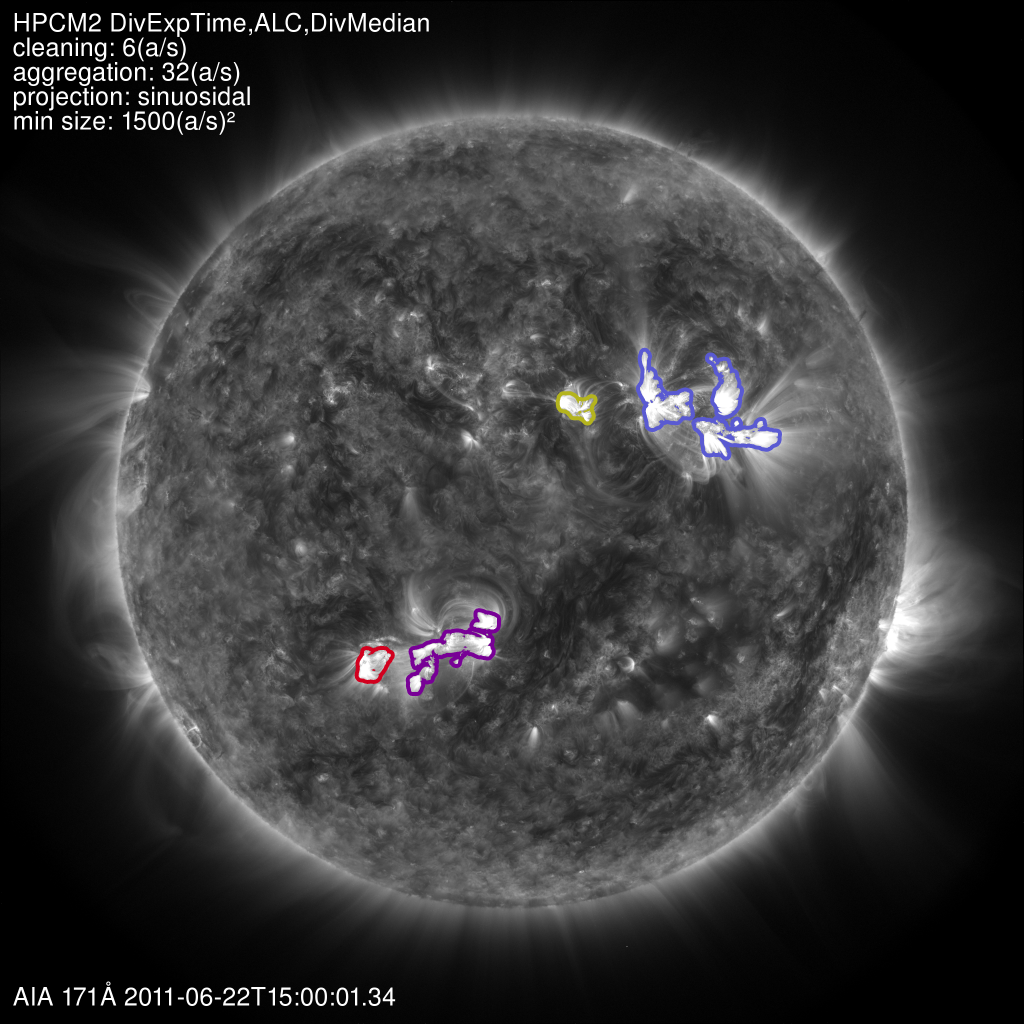}
&
\includegraphics[height=8cm]{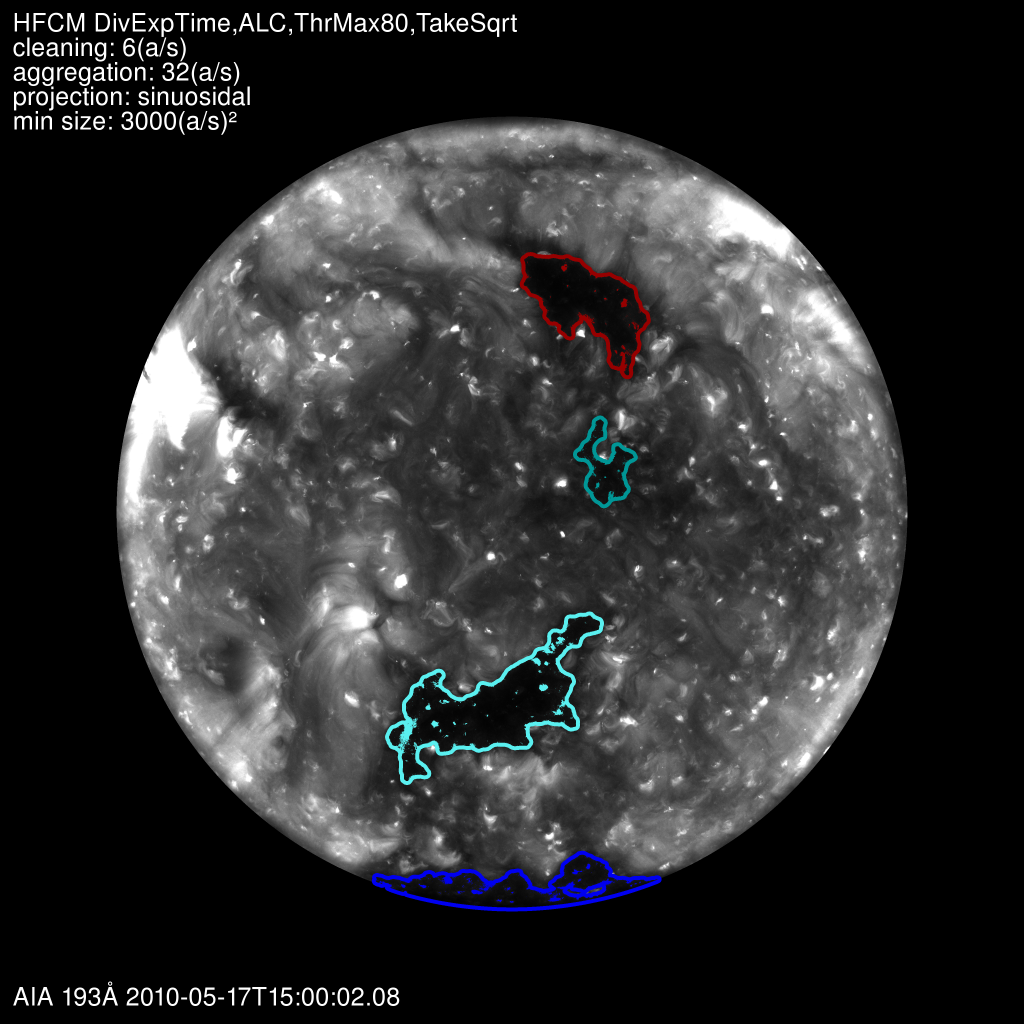}\\
(a) Overlay of AR map on AIA 171 $\AA$ & (b) Overlay  of CH map on AIA 193 $\AA$
\end{tabular}
\caption{\label{F:overlayMap} 
(a) Overlay of a AR map created using AIA 171 and 193 $\AA$ onto the corresponding AIA 171$\AA$  image taken on 20110622 at 15:00
(b) Overlay of CH map created using AIA 193 $\AA$ onto the corresponding AIA 193 $\AA$ image taken on 20100517 at 15:00}
\end{center}
\end{figure*}

\begin{figure*}[htpb]
\vspace*{2mm}
\begin{center}
\includegraphics[height=8cm]{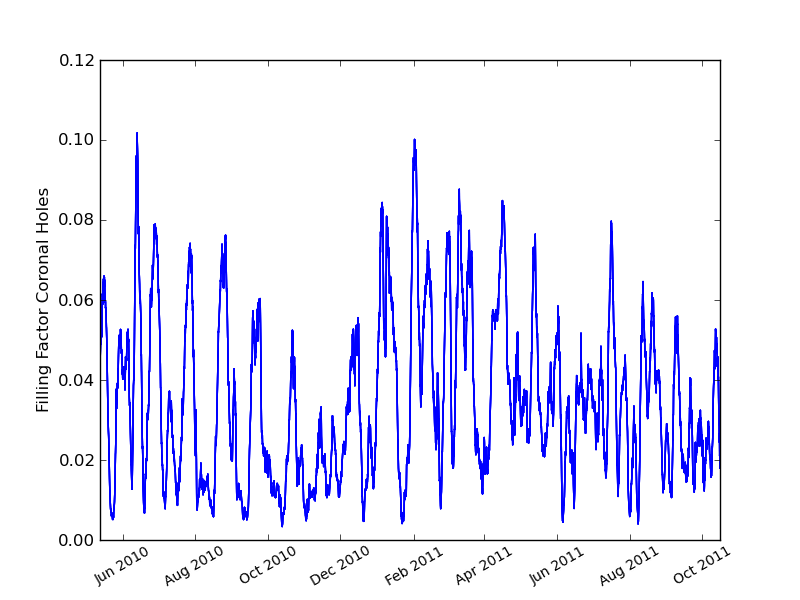}
\caption{\label{F:FF_CH_June2010_Oct2011} Coronal Holes Filling factor extracted from SPOCA results, from 1st June 2010 till 31 October 2011. }
 \end{center}
 \end{figure*}

We now illustrate the results from SPoCA using  SDO images. 
Figure~\ref{F:overlayMap} shows an example of overlays of  AR and CH maps  onto the corresponding AIA images.
Figure~\ref{F:FF_CH_June2010_Oct2011} shows an example of statistics given by the SPOCA algorithm. 
for the period June 2010 till October 2011.

%%%%%%%%%%%%%%%%%%%%ù COMMENTING %%%%%%%%%%%%%%%%%%%%
\begin{comment} 

\deleteopen
(to be  modified)
Their size is however 16 larger than EIT images. To avoid prohibitive computation time,  iterations of the fuzzy clustering are done on the histogram of the pre-processed images rather than on the pre-processed images. A binsize of 0.01 was found to be sufficiently small to produce accurate results. 
\deleteclose

\section{Segmentation of SOHO-EIT archive}
\label{S:eit}
SOHO-EIT images are relatively small $1k \times 1k$, which allows for a fast treatment. The SOHO spacecraft however is situated at the L1-Lagrange point and is thereby exposed to cosmic ray hits (CRH) and to proton events, which deteriorate the quality of the image. The spatial regularization scheme proposed in~\cite{2009A&A...505..361B} is thus necessary to treat EIT images.

Note that the SDO mission on the other hand operates in geo-synchroneous orbit. The SDO-AIA images are therefore less contaminated by CRH, and do not suffer from proton events.

\section{Segmentation of PROBA2-SWAP images }
\label{S:swap}

\section{Segmentation of STEREO-EUVI}
\label{S:euvi}
[What about STEREO-EUVI: do we also use the spatial regularization scheme of SPOCA? Do we need sursegmentation]
%%%%%%%%%%%%%%%%%%%%%%%%%%%%%%%%%%%%%%

%%%%%%%%%%%%%%%%
\end{comment}

\section{Conclusion}

We describe here the algorithm implemented int eh SPOCA-AR and SPOCA-CH modules of the HEK. This provides catalogs or Active Regions and Coronal Holes containing properties such as localisation (through a bounding box or a chain code), area, moments of intensities. 
Contours can also be visualized through \emph{isolsearch} interface or within the \emph{JHelioviewer} visualization software.

These catalogs have many different applications, one of them being the determination of filling factors for CH, AR, QS, to be included into (semi-)empirical models of irradiance.

In future reseach, we plan on improving criteria for distinguishing between filaments and coronal holes. At present the distinction is based on area (we retain only CH larger than $3000$ arcsec$^2$) and duration (we retain only CH older than 3 days), as our study on the year 2010-2011 shows this extract CH candidates having appropriate magnetic properties. As we reach solar maxima however, such distinction between filaments channel and coronal holes might become more problematic.

\appendix
\section{Constraints on regularization parameter}
\label{A:constraint_eta}

All algorithms based on Possibilistic C-means strongly depends on the choice of the regularization parameter $\eta$. Various more or less elaborated procedures have been proposed in the literature, see for example~\cite{Krishnapuram96}. An intuitive choice is to compute $\eta_i$ as the intra-class dispersion, as in equation~(\ref{E:eta}). Problems arise however when the underlying classes have a widely different intra-class variance. For example, the highly variable AR class on EUV images may include in the final segmentation the darkest part of what should be the Coronal Hole class.
 
 To understand this phenomena, consider two classes $C_1$ and $C_2$ with centers $b_1$ and $b_2$. , Suppose $b_{1p} < b_{2p}, \forall p = 1, \dots, P$, where $P$ is the dimension of the feature space. We show that under certain circumstances, $u_{2j}$ can exceed $u_{1j}$ for values of $x_j$ where $x_{jp} < b_{1p}, \forall p = 1, \dots, P$.
 
Let us first determine the locus of points $x_j$ where $u_{1j} = u_{2j}$. By equation~(\ref{E:u_ij}) we found that 
$u_{1j} = u_{2j}$ if and only if
\begin{equation}
\frac{d^2(x_j,b_1)}{\eta_1} = \frac{d^2(x_j,b_2)}{\eta_2}.
\end{equation}
If $\eta_1 = \eta_2$, the solution is a hyperplane through the middle of $b_1$ and $b_2$, and there are no undesired effects.
In case $\eta_1 \neq \eta_2$, and for the Euclidean distance, the above is equivalent to saying that $x_j$ lies on a circle with center
\begin{equation}
c = \frac{\eta_2 b_1 - \eta_1 b_2}{\eta_2 - \eta_1}
\end{equation}
and radius
\begin{equation}
r = \frac{\sqrt{\eta_1 \eta_2}}{|\eta_2 - \eta_1|} \cdot d(b_1,b_2).
\end{equation}
We consider only the case
$\eta_1 < \eta_2$, as this  is the case for the classes CH, QS and AR as they arise in EUV images. In this case, $c$ will lie relatively close to $b_1$, at the ``opposite side'' from $b_2$. All points $x_j$ inside the circle will satisfy $u_{1j} > u_{2j}$, and hence belong to class 1. All points $x_j$ outside the circle will satisfy $u_{2j} > u_{1j}$, and hence will be classified as belonging to class 2. This is unwanted behavior for those points for which some $x_{jp} < b_{1p}$. In order to avoid this situation, we can select $\eta$-values in such a way that the circle center $c$ lies below all coordinate axes. The result is that for all points $x_j$ in the circle, all {\it positive\/} points ``below'' it are also in the circle. Hence whenever a point $x_j$ belongs to class 1, all points ``below'' $x_j$ also belong to class 1. So we want $c_p < 0, \forall p=1, \dots, {\rm dim}$, which is equivalent to the following conditions on $\eta_1$ and $\eta_2$:
\begin{equation}
\frac{\eta_2}{\eta_1} < \frac{b_{2p}}{b_{1p}}, \forall p = 1, \dots, {\rm dim}
\end{equation}

\section{Computation of chain code}
\label{A:chain_code}

Chain coding aims at representing the boundary of an object in digitized images. It is based on the idea of following the outer edge of the object and storing the direction when travelling along the boundary~\citep{freeman61}.
In the SPOCA-suite, we use a representation of the chain code with eight directions, as is commonly done in the literature. Such represenation is however of the same length as the perimeter of the object under consideration, which in many cases is too long.
In a second step, we thus find a polygonal approximation to the perimeter that has a maximal number of edges, and for which the distance from any point in the perimeter to the polygon does not exceed a given accuracy.

The algorithm proceeds as a recursive refinement. The main axis of the contour is
first extracted, providing the first two vertices. Each polygon edge is then recursively split by introducing a
new vertex at the most distant associated contour point, until the desired accuracy is reached. More precisely, the algorithm runs as follows: 
 \begin{enumerate}
\item Initialize the polygon with points $p_1$ and $p_2$ of the perimeter that are furthest away from each other.
\item Let $i=3$ 
\item  For each segment in the polygon find the point on the perimeter that have the furthest distance to the polygonal line-segment. If this distance is larger than a threshold mark the point with a label $p_i$
\item Renumber the points so that they are consecutive
\item Increase $i$
\item If no points have been added then finish, otherwise go back to step 3.
\end{enumerate}

Within the HEK, a maximal number of 100 points is considered to be sufficient for a reasonable approximation of CH boundaries.

\begin{comment}

From the code of Benjamin: 
To extract the chain code of a connected component, we first search the first pixel on the external boundary.
Then we list all the points along the boundary starting from that first pixel.
Once The list is made, it is reduced by trying to find the most relevant points.
First we take the firstpixel point and it's furthest point in the list, and add them to the chain code.
Then we search the point that is the furthest from the line passing by each pair of consecutive point in the chain code, and add it to the chain code.
We repeat that last step until we have enough points

param image The image to use for the computation of the chaincode
param max-points The maximal number of chaincode points to list.
param max-deviation The maximal deviation of the chaincode curve between 2 points, in arcsec.
\end{comment}

%---- Acknowledgements 
\section*{Acknowledgements}
\label{S:acknowledgements}

The research leading to these results has received funding from the European Commission's Seventh Framework Programme (FP7/2007-2013) under the grant agreement n0 263506 (AFFECTS project).
Funding of CV, VD, and BM by the Belgian Federal Science Policy Office (BELSPO) through the ESA/PRODEX Telescience and SIDC Exploitation programs is hereby acknowledged. CV was also supported by the Solar-Terrestrial Center of Excellence. 
We acknowledge support from ISSI through funding for the International Team on SDO datamining and exploitation in Europe (Leader: V. Delouille).

%%%%%%%%%%%% Bibliography %%%%%%%%%%%%%%%%

\bibliographystyle{aa}
\bibliography{biblio,issi_database}

%%%%%%%%%%% FIGURES %%%%%%%%%%%%%%%%%%%%%

 \begin{figure*}[htpb]
 \vspace*{2mm}
 \begin{center}
 \begin{tabular}{cc}
 \includegraphics[height=6cm]{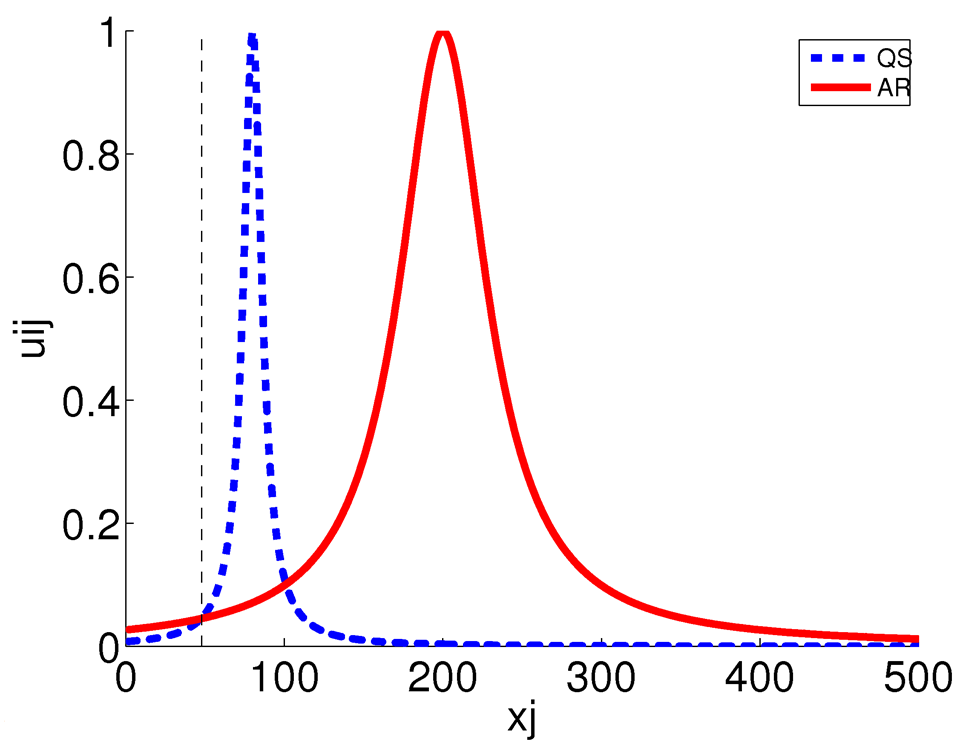}
 &
 \includegraphics[height=6cm]{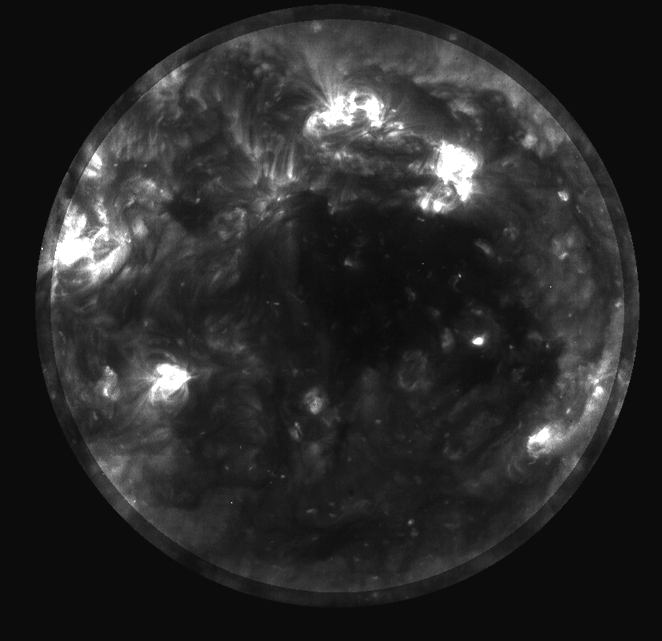}\\
 (a) membership function for AR and QS & (b) EIT 195 $\AA$ image\\
 \includegraphics[width=6cm]{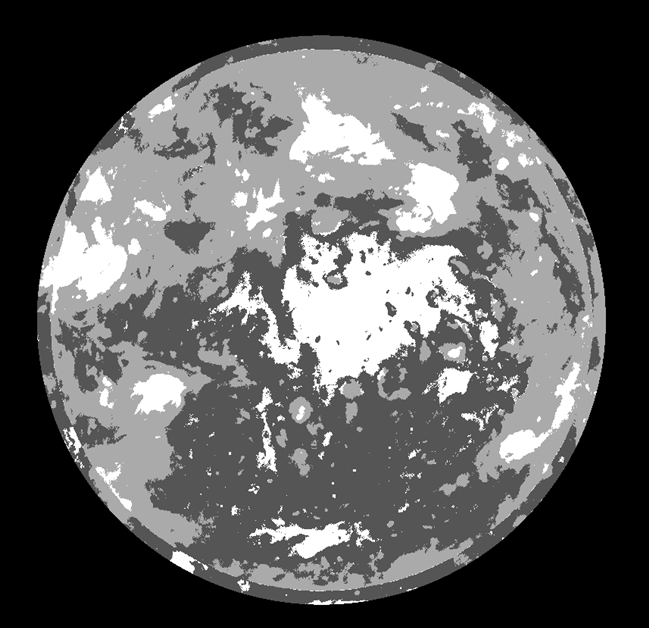}
 &
 \includegraphics[width=6cm]{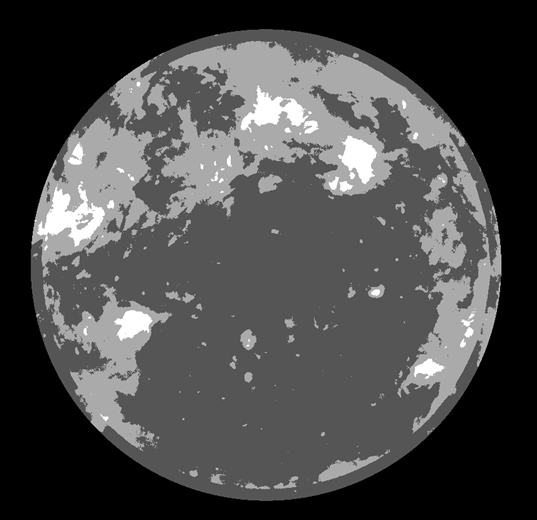}\\
 (c) Original segmented map  & (d) Segmented map with constraints
 \end{tabular}
 \caption{\label{F:eta-condition} 
 (a) Illustration of membership functions $u_{ij}$ for QS and AR descriptors $x_j$. Because of the larger spread in intensity values of the AR, small values of $x_j$ may have a larger AR membership than QS membership. (b) EIT 195 $\AA$ from January 1, 2000 with correction of limb brightness (c) Segmentation using original PCM algorithm: darkest part are classified as Active regions (d) Segmentation using PCM with constraints on $\eta$-values}
 \end{center}
 \end{figure*}

\end{document}